\documentclass[reprint,amsmath,amssymb,aps,prb,superscriptaddress]{revtex4-2}
\usepackage{graphicx}
\usepackage{dcolumn}
\usepackage{bm}
\usepackage{color}
\usepackage{blindtext}
\usepackage{float}
\usepackage{hyperref}
\usepackage{amsmath}
\usepackage{enumerate}
\usepackage{changes}
\hypersetup{colorlinks=true, allcolors=blue}
\begin{document}
\preprint{APS/PRB}
\title{Band Gap Engineering of Nitrogen-Doped Monolayer WSe$_2$ Superlattice and its application to Field Effect Transistor
}%

\author{Yi-Cheng Lo}%
\affiliation{%
Department of Electrophysics, National Yang Ming Chiao Tung University, 1001 Daxue Rd., Hsinchu City 300093, Taiwan
}%
\author{Liao Jia Wang}%
\affiliation{%
Department of Electrophysics, National Yang Ming Chiao Tung University, 1001 Daxue Rd., Hsinchu City 300093, Taiwan
}%

\author{Yu-Chang Chen}
\email{Corresponding author: yuchangchen@nycu.edu.tw}
\affiliation{%
Department of Electrophysics, National Yang Ming Chiao Tung University, 1001 Daxue Rd., Hsinchu City 300093, Taiwan
}%
\affiliation{Center for Theoretical and Computational Physics, National Yang Ming Chiao Tung University.}


\date{\today}

\begin{abstract}
We systematically investigate the electronic structures of pristine monolayer WSe$_2$ and WSe$_2$ superlattices with periodic nitrogen substitution. Unlike random doping, which often introduces in-gap impurity states, periodic nitrogen doping primarily modulates the band gap, thereby facilitating effective band gap engineering for electronic and optoelectronic applications. The gap narrows monotonically with increasing dopant density (pristine $>$ 8-row $>$ 6-row $>$ 4-row), directly influencing device switching. We also evaluate the FET performance of nanojunctions created by these configurations by examining the contour plot of current density as a function of temperature and gate voltage, which quantifies how bandgap engineering affects switching characteristics. Our calculations clarify the classical-quantum crossover in sub-10 nm 2D FETs: as $T$ rises, $J$ approaches the thermionic current; as $T$ falls, quantum tunneling dominates, and the steep energy dependence of $\tau(E)$ may break the classical limit of subthreshold swing imposed by the Boltzmann tyranny. The optimal gating range ($V_g^\mathrm{ON}$, $V_g^\mathrm{OFF}$) is investigated for each temperature, insensitive to temperature in the high-temperature regime, confirming the good thermal stability of the FET devices. A comparison study demonstrates that the 4-row structure, with excessively large $J_\mathrm{OFF}$, severely low ON/OFF ratio, and restricted operation range, is inappropriate for realistic FET applications. The pristine structure has the highest performance across all measures, but its high $V_g^\mathrm{OFF}$ ($\sim$1.1 V) makes it less practical, since such a large threshold voltage may promote time-dependent dielectric breakdown (TDDB) of the oxide layer, reducing device dependability.  The 6-row and 8-row structures are slightly inferior to the pristine in terms of performance, but exhibit more favorable $V_g^\mathrm{OFF}$ values ($\sim$0.75 V), achieving a balance between reasonable threshold voltage and stable operation range, making them more promising candidates for future FET integration.
\end{abstract}

\maketitle


\section{Introduction}\label{sec:intro}

The relentless pursuit of miniaturization in transistor technology has become a defining challenge for the semiconductor industry, serving as a pivotal catalyst for advancements in computing capabilities. ~\cite{Radosavljevic_IEEE_2022}. The ability to fabricate smaller transistors allows for a greater number of devices to be integrated onto a single chip, thereby amplifying computational power while simultaneously enhancing performance metrics such as switching speed and power efficiency.~\cite{Keyes_IEEE_2006}.  However, as the industry approaches the physical limits of silicon-based transistors, significant hurdles arise, including diminished carrier mobility and increased leakage currents due to quantum tunneling effects as channel dimensions shrink to sub-10 nm scales.~\cite{Natelson_PhysicsWorld_2009} 

In response to these challenges, two-dimensional (2D) transition metal dichalcogenides (TMDs) have emerged as highly promising alternatives, boasting superior electronic and structural characteristics that facilitate improved carrier mobility and integration into three-dimensional architectures~\cite{Duan_Chem_Rev_2024,mezeli_nat_rev_matt_2017}. The carrier mobility in transition metal dichalcogenides (TMDs) has been improved to exceed that of silicon \cite{Akinwande_Nature_2019}, and further enhancement may be achieved through lattice distortion \cite{Ng_Nat_Elct_2022}. Two-dimensional (2D) materials can be integrated into three-dimensional (3D) architectures to further increase transistor density on a chip~\cite{Lee_NanoLett_2024}.

The performance of field-effect transistors (FETs) is typically characterized by key metrics such as the OFF current, ON/OFF current ratio, and subthreshold swing (S.S.). In the silicon-based semiconductor industry, scaling down the channel length of transistors from the micrometer scale to sub-10 nm introduces significant challenges related to material properties and changes in transport mechanisms~\cite{YC_ACSNano_2025,YC_FET_2025}. Specifically: (1) the carrier mobility of silicon as a channel material decreases sharply; and (2) strong quantum tunneling effects lead to increased leakage currents, thereby degrading the OFF current. ~\cite{Ieong_Science_2004,Tze-chiang_IEEE_EXPLORE_2009}

Most 2D TMD-based junctions are fabricated using a top-contact geometry with channel lengths exceeding the electron–phonon mean free path~\cite{Radisavljevic_Nat_nanoten_2011,Najmaei_ACSNano_2014,Liu_NanoLett_2016}, which leads to higher resistance and increased heat dissipation through Joule heating. In addition, van der Waals interactions at the interface between TMD materials and metal electrodes often result in substantial contact resistance due to the formation of a Schottky barrier~\cite{Novoselov_Science_2016,Somvanshi_PRB_2017,Wu_Nat_Rev_Mat_2023}. This elevated contact resistance reduces the ON current and weakens the overall transistor signal. Consequently, optimizing the metal–TMD interface to minimize contact resistance is critical for improving device performance\cite{Boehm_NanoLett_2023}. Significant efforts, such as doping~\cite{Khalil_ACS_Appl_Mat_Int_2015,Jiang_nat_elect_2024,Sahoo_AdvFuncMatt_2025}, have been devoted to addressing this issue. 

Another important characteristic of two-dimensional (2D) field-effect transistors (FETs) is their band gap. Tailoring the band gap of transition-metal dichalcogenide (TMD) semiconductors is essential for controlling their electrical and optical properties in specific applications. Band gap engineering can be achieved through doping~\cite{Yoo_nanomaterials_2021}, strain application~\cite{Bolan_NanoLett_d2024}, or layer stacking~\cite{Chaves_npj2D2020}. For example, nitrogen substitutional doping in WSe$_2$ has been reported to modulate its electronic properties; however, the underlying mechanisms governing its influence on transport behavior and transistor performance remains insufficiently understood~\cite{NitrogenDopingBandGapTuning}.

To address this challenge, the present study proposes the use of periodic doping in 2D TMD materials as an effective strategy for band gap engineering. Controlled periodic doping may mitigate the formation of disordered impurity states and transform a monolayer TMD into a superlattice—a man-made crystal with unique electronic properties arising from its periodic structure. Compared with random impurities, the resulting electronic structures are expected to be cleaner and more efficient in modulating the band gap. 

\begin{figure*}
\centering
\includegraphics[width=0.8\textwidth]{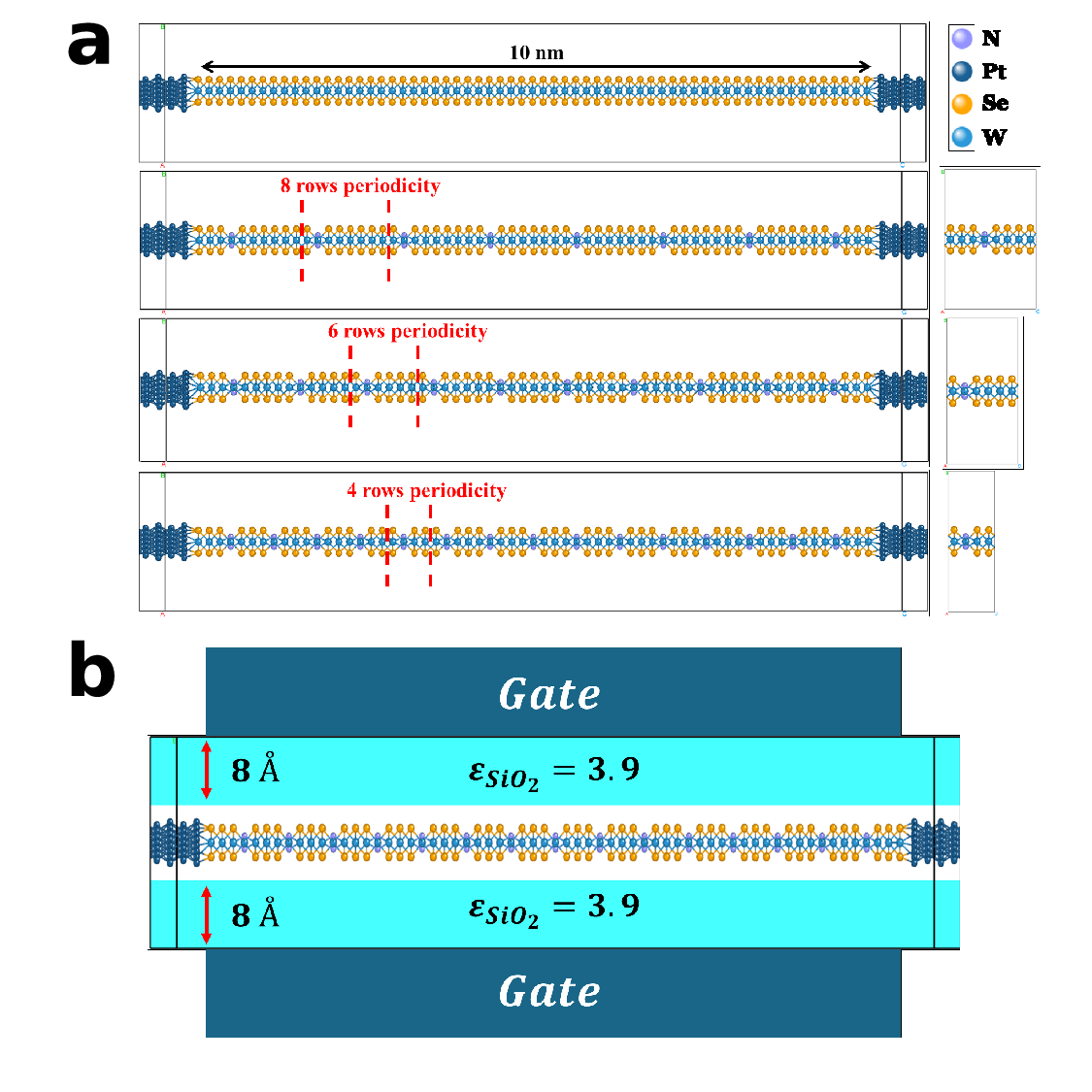} 
\caption{a) Schematic of the four representative junction configurations: pristine WSe$_2$ and nitrogen-doped WSe$_2$ superlattices with 8-row, 6-row, and 4-row doping periodicities (from top to bottom), bridged between two platinum electrodes to form nanojunctions with a channel length of around 10 nm.
b) Schematic of the gate architecture for the four junctions configured as field-effect transistors (FETs) in DFT–NEGF (NanoDCAL) simulations. The gate is modeled with an equivalent oxide thickness (EOT) of 8 Å and a dielectric constant of 3.9.
}
\label{fig:scheme}
\end{figure*}

Motivated by these considerations, this study proposes periodic nitrogen doping in monolayer WSe$_2$ as an effective superlattice-based route to band gap engineering. We posit that such a strategy can achieve substantial band gap modulation while simultaneously mitigating contact resistance and enhancing electronic transport properties. Using first-principles calculations within the framework of density functional theory (DFT), we investigate the electronic structures of WSe$_2$ superlattices formed by controlled nitrogen substitution with periodicities of $n$ rows. Four representative configurations are examined: pristine WSe$_2$ and superlattices with 8-row, 6-row, and 4-row nitrogen-doping periodicities. These monolayer configurations are embedded between platinum electrodes to form Pt–WSe$_2$–Pt nanojunctions with channel lengths of approximately 10 nm, as illustrated in Figure~\ref{fig:scheme}$\bf{a}$. A gate architecture, depicted in Figure~\ref{fig:scheme}$\bf{b}$, is employed for all four configurations to evaluate the transmission coefficient $\tau(E)$ using DFT combined with the nonequilibrium Green’s function (NEGF) formalism. The equivalent oxide thickness (EOT) of the gate is modeled using a dielectric layer with a relative permittivity of 3.9 and a thickness of 8 \AA~\cite{FET-AlN}. 

In addition to electronic structure and band gap analysis, we systematically investigate the impact of periodic nitrogen doping in monolayer WSe$_2$ on the performance of two-dimensional FET devices. Key device metrics are evaluated, including ON/OFF current ratios, subthreshold swing, and the optimal operational window, over a temperature range of $100–500$ K and gate voltages from $–1.0$ to $1.5$ V.

\section{Theory and Methods}\label{sec:theory}

\subsection{VASP}

The Vienna Ab-initio Simulation Package (VASP) was used to calculate and optimize the electronic structures of nitrogen-doped WSe$_2$ superlattice monolayers as well as the geometries of the platinum electrodes, nitrogen-doped WSe$_2$ superlattice monolayer, and the combined Pt–WSe$_2$–Pt nanojunctions, as shown in Figure~\ref{fig:scheme}$\bf{a}$. VASP utilized the full-potential projected augmented wave method with a plane-wave basis to address the Kohn-Sham problem effectively. \cite{VASP1, VASP2, VASP3, VASP4}  In these density functional theory (DFT) computations, the Perdew-Burke-Ernzerhof functional (PBE), a variant of Generalized Gradient Approximation (GGA), is utilized to address many-body effects within the effective single-particle framework.  All calculations employed a grid size of $0.016$~\AA$^{-1}$ in reciprocal space and a plane-wave energy cutoff of $400$~eV.  The criterion for terminating the electronic self-consistent iterations was established at $1.0\times 10^{-4}$ eV.

\subsection{NanoDCAL }\label{subsec:NanoDCAL}

NanoDCAL (Nanoacademic Device Calculator) is used to compute the transmission functions, $\tau(E)$, in consideration of the source-drain voltage $V_\text{ds}=50$ mV and the gate voltage $V_{\text{g}}$. NanoDCAL  performs self-consistent calculations using the Keldysh nonequilibrium Green’s function formalism in conjunction with a linear combination of atomic orbitals (LCAO), within the framework of density functional theory (DFT)~\cite{NanoDCAL1, NanoDCAL2, Keldysh}. Troullier-Martins norm-conserving pseudopotentials were utilized to simulate electron-ionic core interactions. Double-$\zeta$ polarized basis sets were utilized to examine elemental valence electrons. The PBE-GGA exchange-correlation functional was chosen.~\cite{PBE, GGA} The equivalent energy cutoff for the grid density was set to $100$~Hartree. The Brillouin zone in reciprocal space was sampled using a $6 \times 1 \times 100$ k-point grid for the electrodes and a $6 \times 1 \times 1$ grid for the central scattering region. The $k$-point grids used for calculating the transmission coefficient and current were $6 \times 1 \times 1$ and $1 \times 1 \times 100$, respectively.

\subsection{\label{sec:asymptotic}Asymptotic behavior of $\tau(E)$ and correspondence principle}

As shown in Ref.~\cite{EMT-PW} and \cite{YC_FET_2025}, the Landauer formula can be derived from the classical expression $J = n e v$ as follows:
\begin{equation}
J_z=\frac{2e}{h} \int  [f^{\text{R}}(E,T)-f^{\text{L}}(E,T)] \tau (E) dE,
\label{eq:Landuaer2}
\end{equation}
where the transmission coefficient is
\begin{equation}
\tau (E)=\frac{m}{\hbar}  \int_{0}^{\infty} |v_{z}(k_z)| \Theta [E-E_{z}(k_z)] dk_z.
\label{tauE2}
\end{equation}
In the classical limit, the transmission coefficient becomes
\begin{equation}
\begin{aligned}
\tau (E)  ^{\text{asymp.}} \approx 
 \frac{m}{2 \pi \hbar^{2}} (E-\chi) \Theta (E-\chi) .
\end{aligned}
\label{tauE3}
\end{equation}
The current density derived from the Landauer formula approaches the thermionic emission current defined by Richardson's law in the classical limit, where $W = \chi - E_F \gg k_B T$ and the Fermi–Dirac distribution $f(E, T)$ reduces to the Boltzmann distribution:
\begin{equation}
\begin{aligned}
J_z &=\frac{2e}{h} \int  [f(E,T)] \tau(E) ^{\text{asymp.}} dE  \\
 &= \left(\frac{e m k_{B}^{2} }{2 \pi^{2} \hbar^{3}} \right ) T^{2} e^{\frac{-W}{k_B T}}, 
\end{aligned}
\label{eq:Richadson}
\end{equation}
where $\left(\frac{e m k_{B}^{2} }{2 \pi^{2} \hbar^{3}} \right )$ is the Richardson constant. This convergence is the result of the correspondence principle in quantum mechanics, which guarantees that quantum results revert to classical behavior in the appropriate limit.

\subsection{Subthreshold swing and its classical limit of Boltzmann Tyranny}\label{subsec:current_and_SS}

The subthreshold swing (S.S.) of a FET quantifies its efficiency in modulating the current density. It represents the required change in gate voltage to increase the output current density by one order of magnitude and is defined as
\begin{equation}
S.S.(T,V_g,V_{ds}) \equiv \ln(10) \left\{ \frac{1}{J(T,V_{ds},V_g)} \frac{dJ(T,V_g,V_{ds})}{dV_g} \right\}^{-1}.
\label{eq:SS}
\end{equation}
Alternatively, within the effective gate model, the subthreshold swing can be expressed as
\begin{equation}
\begin{aligned}
& S.S.(T,V_g,V_{ds}) = \left[ \ln(10) \left ( k_{B} T/e\right ) \right ] \times   \\
& \frac{ \int_{-\infty}^{\infty} \left [f^R(T,V_g,V_{ds})-f^L(T,V_g) \right ]\tau(E) dE }{\int_{-\infty}^{\infty} \left \{  \operatorname{sech}^2 \left [  \frac{E-\mu_R(V_g,V_{ds})}{2 k_B T}    \right ]  
- \operatorname{sech}^2 \left [  \frac{E-\mu_L(V_g)}{2 k_B T}    \right ]    \right \}\tau(E) dE },
\end{aligned}
\label{eq:SS_VG}
\end{equation}
where $\left[ \ln(10), (k_B T / e) \right]$ represents the classical limit of the subthreshold swing in field-effect transistors, commonly referred to as the Boltzmann tyranny.

When the gate voltage $V_g$ moves the chemical potential $\mu(V_g)$ to fall within the transmission band gap region, a competition emerges between quantum tunneling and thermionic emission currents. In scenarios where the thermionic emission current significantly surpasses the electron transport mechanism, leading to a negligible quantum tunneling current, the subthreshold swing converges towards the following limit:
\begin{equation} \label{eq:SSapproximation}
S.S. \rightarrow \ln(10) (k_B T / e).
\end{equation}
which corresponds to the Boltzmann tyranny.


\section{Results and Discussion}\label{sec:results}

\begin{figure*} [ht] 
\includegraphics[width=1.0\linewidth]{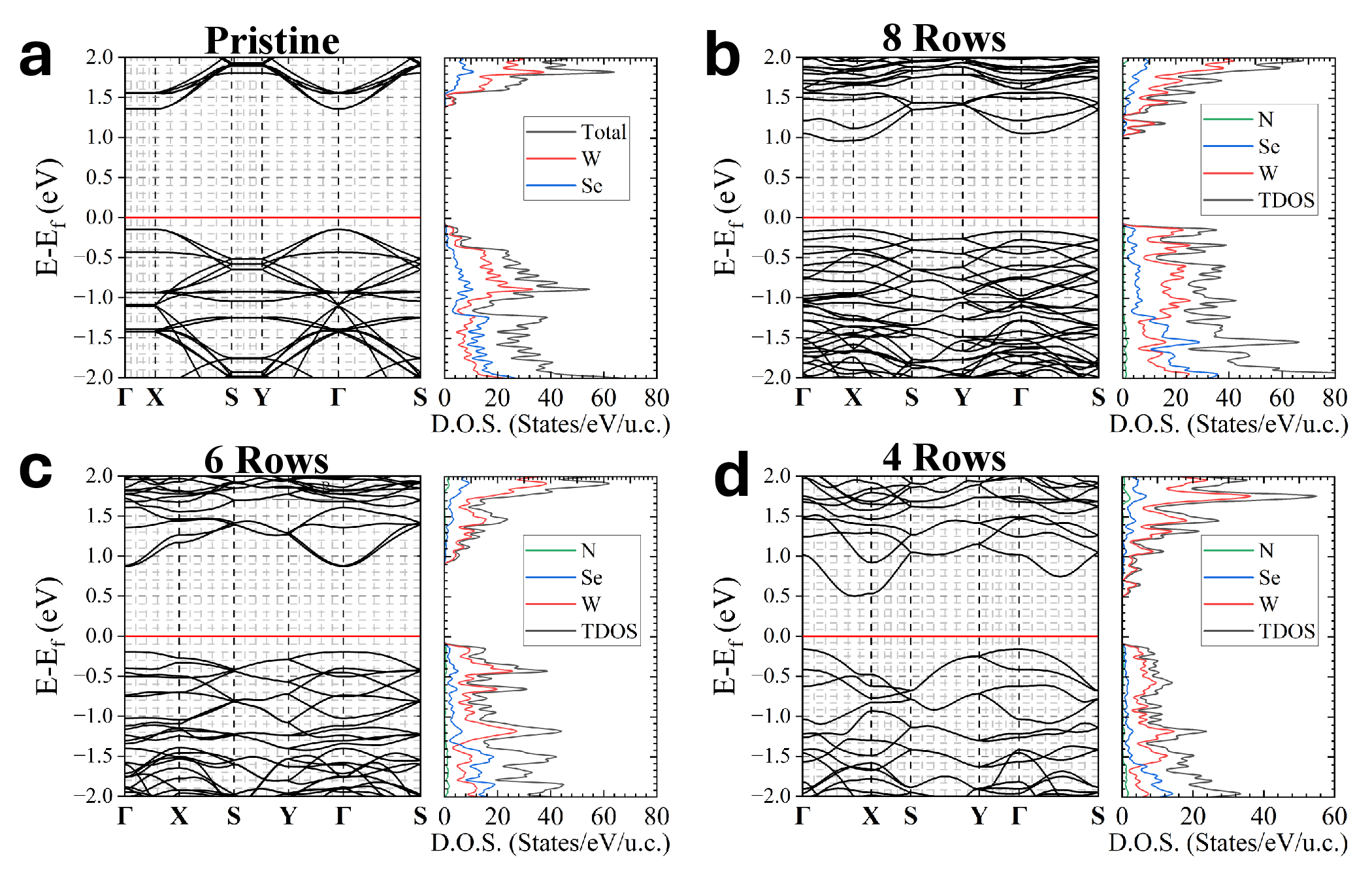}
\caption{
Right panels: band structures. Left panels: total density of states (DOS; black) and atom-projected DOS (PDOS) on W (red), Se (blue), and N (green) for a) pristine WSe$_2$ monolayer and nitrogen-doped superlattices with b) 8-row, c) 6-row, and d) 4-row periodicity. The corresponding band gaps are a) 1.50 eV, b) 1.10 eV, c) 1.07 eV, and d) 0.66 eV.
}
\label{fig:Band}
\end{figure*}

Density Functional Theory (DFT) using VASP was employed to compute the electronic structures of the pristine WSe$_2$ monolayers and nitrogen-doped superlattice monolayers with periodicities of 8-row, 6-row, and 4-row shown in  Figure~\ref{fig:scheme}. The band structures and density of states (including the atomic projected density of states (PDOS) on each atom) of the four periodic configurations are displayed in Figure~\ref{fig:Band}. 
Periodic (regular) nitrogen doping chiefly tunes the band gap, rather than introducing in-gap impurity states as random doping often does. The total density of states (DOS) reveals a clear trend of bandgap reduction with increasing doping concentration. Consequently, nitrogen-doped WSe$_2$ superlattices retain a clean crystalline character while enabling effective band-gap engineering. Furthermore, atomic projected density of states (PDOS) analysis indicates that, across all four configurations, the states near the valence-band maximum (VBM) and conduction-band minimum (CBM) are predominantly derived from W atoms.

In the pristine (undoped) WSe$_2$ monolayer, the band gap is around 1.50 eV with both the valence-band maximum (VBM) and conduction-band minimum (CBM) located at $\Gamma$ point, indicating a direct gap. As the nitrogen concentration increases, the 8-row superlattice shows a reduced gap of  around 1.10 eV, with VBM and CBM both at $X$ point, revealing a direct band gap. For the 6-row structure, VBM and CBM return to $\Gamma$ point, and the direct  band gap further narrows to around 1.07 eV. At the highest doping (4-row), the CBM shifts to $X$ point while the VBM s at $\Gamma$ point, yielding an indirect gap of 0.66 eV.

\begin{figure} 
\includegraphics[width=\linewidth]{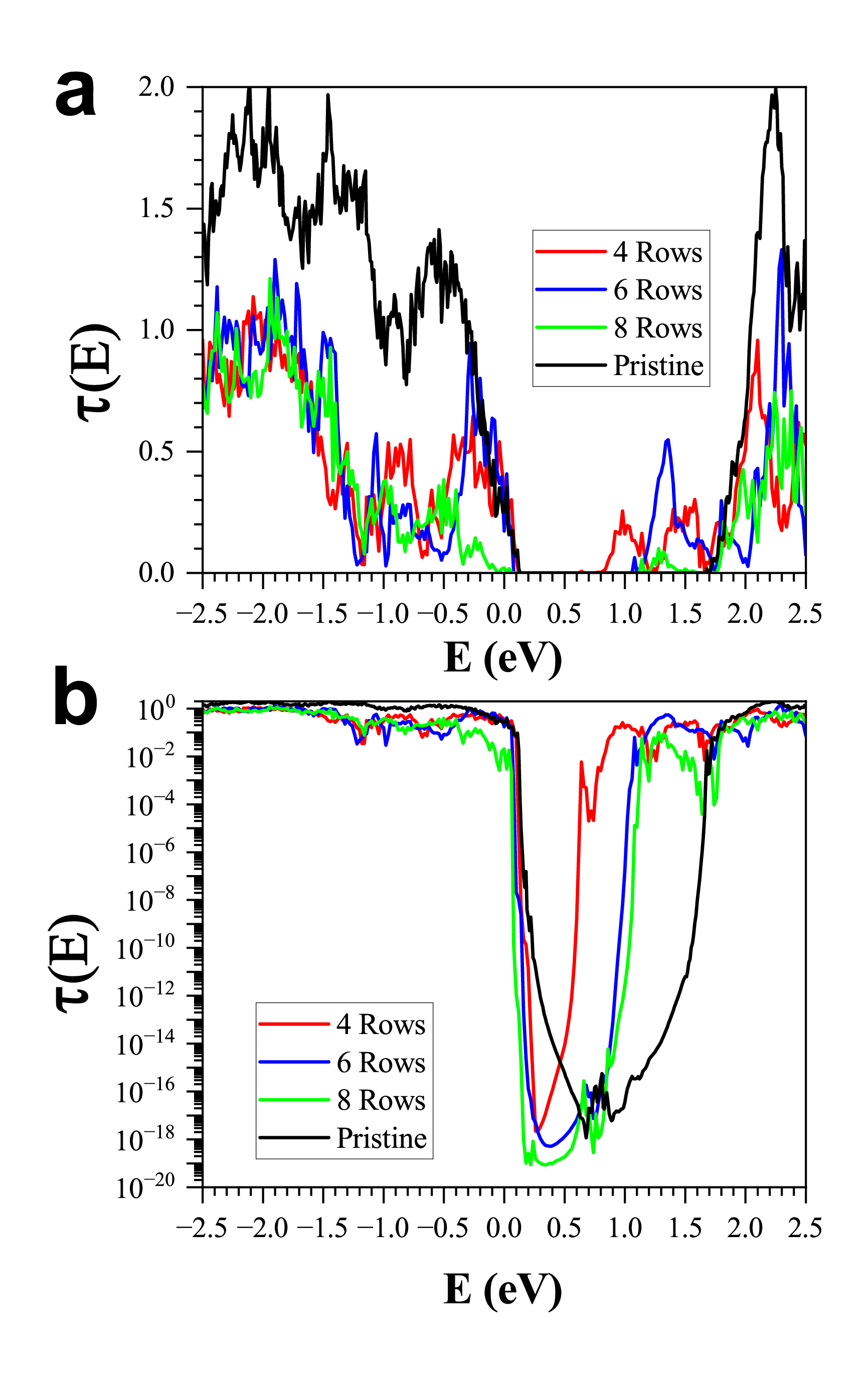}[ht]
\caption{
Transmission spectra for the four configurations depicted in Figure~\ref{fig:scheme}$\bf{a}$ at $V_{ds} = 50$ mV, presented on a) linear and b) logarithmic scales.  The VBM positions  consistent across all instances, but the CBM positions vary considerably, resulting in a decrease in the transmission band gap as nitrogen concentration increases.  The minimum values of $\tau(E)$ exhibit minimal sensitivity to doping, demonstrating just a marginal further suppression with increasing nitrogen concentration.
}
\label{fig:Transmission}
\end{figure}

NanoDCAL is employed to compute the transmission functions, $\tau(E)$, at $V_{ds}=50$ mV. From the transmission spectra of the four structures in Figure~\ref{fig:Transmission}$\bf{a}$, distinct transmission band gaps can be identified, exhibiting clear $p-$type semiconducting behavior. The valence band maximum (VBM) for all four configurations s pinned at approximately the same energy. In contrast, the conduction band minimum (CBM) shifts progressively to lower energies with increasing dopant concentration, leading to a reduction in the band gap. Notably, side peaks in $\tau(E)$ emerge above the CBM as the nitrogen concentration increases. Overall, periodic nitrogen doping promotes the emergence of side peaks that partially fill the pristine structure's  band gap, thereby effectively narrowing the transmission gap, rather than creating localized impurity states in the band gap region as random doping often does. Consequently, regularly nitrogen‐doped monolayer WSe$_2$ forms a man-made crystal, and such a superlattice structure sandwiched between metal electrodes with a gate architecture offers a clean and efficient method for band gap engineering for 2D monolayered field effect transistors. The band gaps of pure and nitrogen-doped monolayer WSe$_2$, along with the nanojunctions they create, are presented in Table \ref{table:bandgap}.


\begin{table}[h!]
\centering
\caption{Band gaps for pristine and periodic nitrogen-doped monolayer WSe$_2$, and the nanojunctions formed by them.}
\begin{tabular}{lcccc}
\toprule 
\textbf{Periodicity} & \textbf{Pristine} & \textbf{8-Row} & \textbf{6-Row} & \textbf{4-Row} \\
   &   (eV) &   (eV) &   (eV) &   (eV) \\
\hline
Super Lattice & 1.5049 & 1.1021 & 1.0687 & 0.6628 \\
Nanojunction & 1.5600 & 1.2050 & 0.9700 & 0.5400 \\
\hline
\end{tabular}
\label{table:bandgap}
\end{table}

The superlattice structure positioned between electrodes creates a potential barrier, resulting in a non-zero transmission coefficient within the band gap region.  The transmission coefficients at $V_{ds}=50$ mV are depicted on a logarithmic scale in Figure~\ref{fig:Transmission}$\bf{b}$.
The transmission spectrum exhibits an increasingly steeper rise near the VBM as the dopant density increases. Nitrogen substitution perturbs the local bonding environment and introduces additional states near the valence band, thereby sharpening the transmission spectrum at the VBM. Under finite bias, these states couple more strongly to the asymmetric broadening functions, which further amplifies the steepness of the transmission edge. Such enhanced steepness may signify a reduced ON–OFF voltage ratio.


Nitrogen substitution perturbs the local bonding and introduces additional states near the valence band, sharpening the features in the transmission spectrum at the VBM. Under finite bias, these states interact more strongly with the asymmetric broadening functions, amplifying the steepness of the transmission edge. This steepness could indicate a smaller on-off voltage ratio. 

\begin{figure} 
\centering
\includegraphics[width=0.5\textwidth]{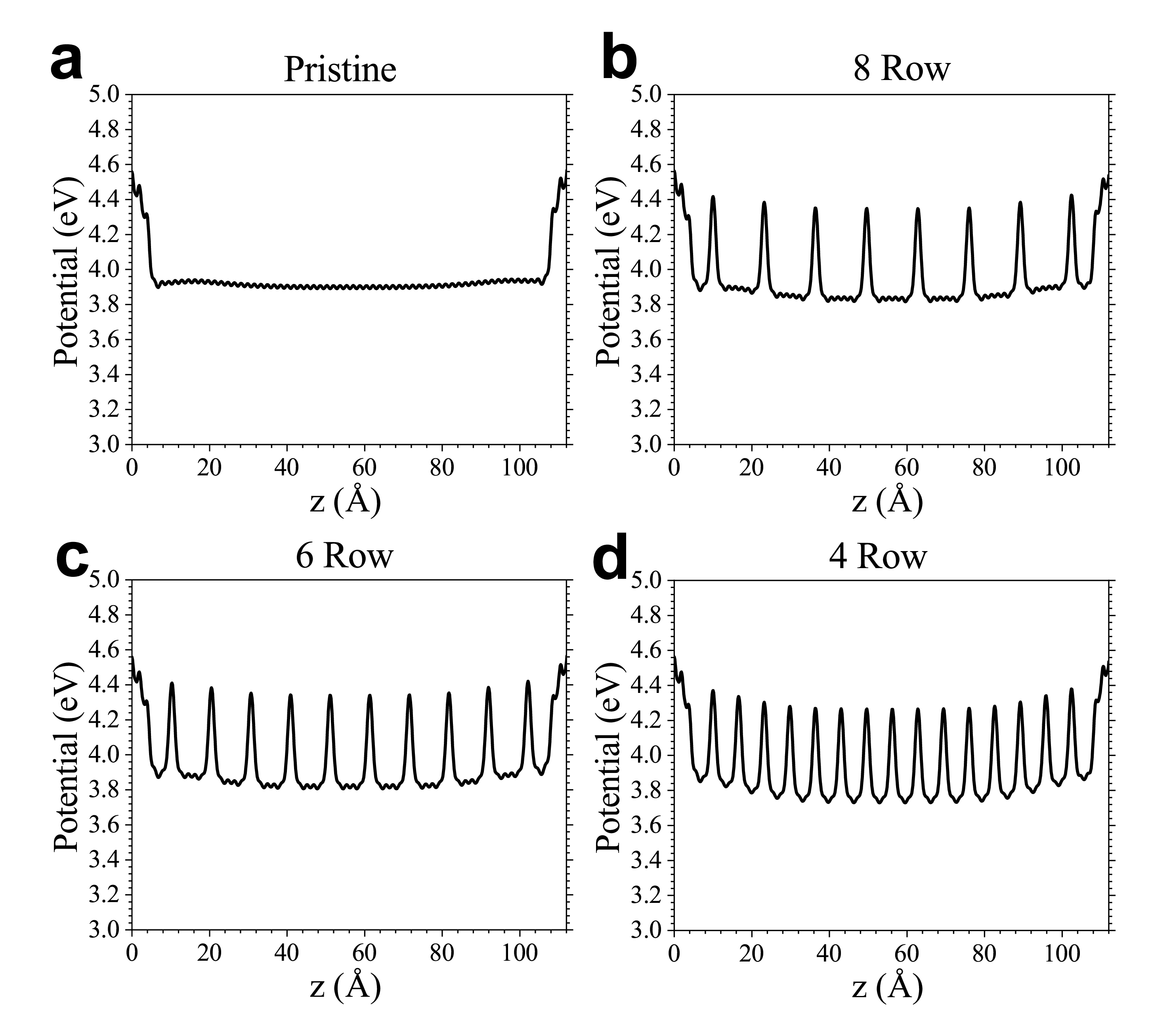}
\caption{ 
Electrostatic potential profiles of the four nanojunction structures: a) pristine, b) 8-row, c) 6-row, and d) 4-row configurations depicted in Figure~\ref{fig:scheme}$\bf{a}$. In the channel region, the pristine junction exhibits a uniform potential landscape, whereas the 8-row, 6-row, and 4-row configurations display equally spaced potential barriers arising from periodic nitrogen doping.
}
\label{fig:Potential}
\end{figure}

Figure.~\ref{fig:Potential} shows the Hartree potential obtained from NanoDCAL. In the pristine configuration, the channel s uniform and free of pronounced modulations, forming only a simple potential barrier. In contrast, periodic nitrogen doping introduces distinct peaks in the electrostatic potential, arising from the regular dopant distribution and averaged over the cross-section perpendicular to the transport direction. Because nitrogen is much more electronegative than selenium, it withdraws electron density from neighboring tungsten and selenium atoms. This redistribution localizes charge and enhances the positive electrostatic potential, effectively generating screened barriers for charge carriers. As a result, the periodic arrangement of nitrogen atoms partitions the channel into a sequence of quantum wells separated by potential barriers, thereby transforming the originally uniform WSe$_2$ channel into an artificial crystal structure.


\begin{figure*} 
\centering
\includegraphics[width=1.0\textwidth]{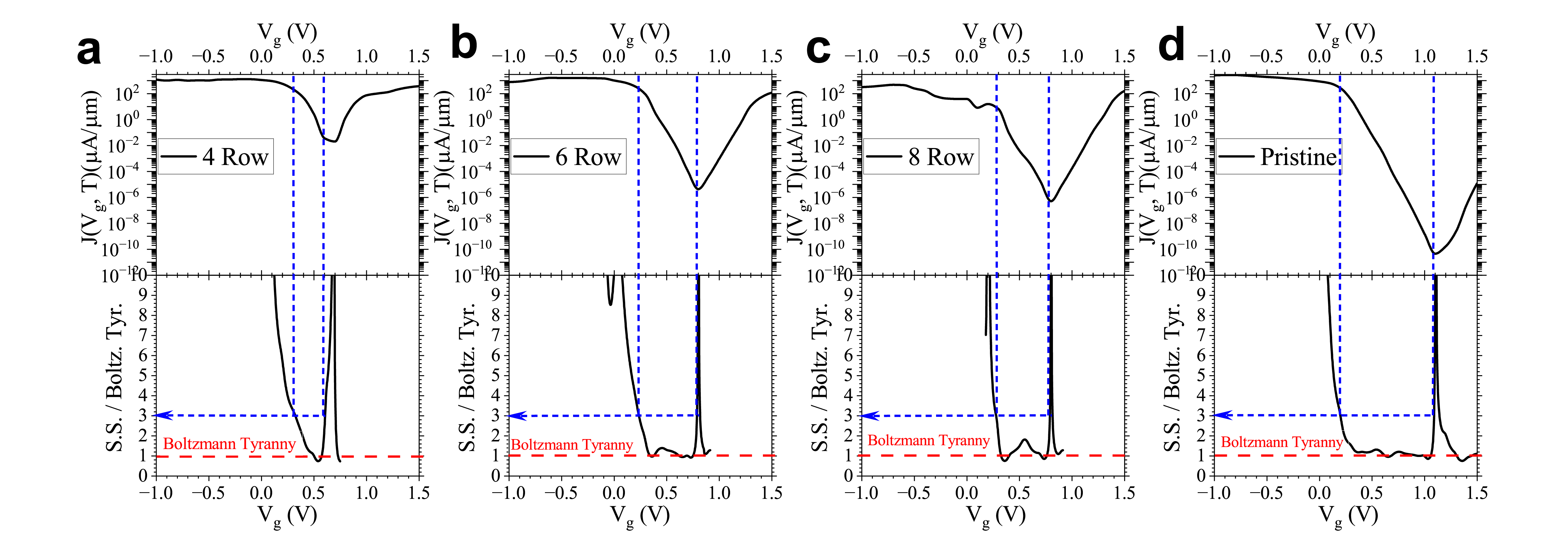}
\caption{
Log($J$)–$V_{\mathrm{g}}$ characteristics of the four structures at room temperature. As $V_{\mathrm{g}}$ increases, a negatively sloped linear region appears, corresponding to the operational range of each structure. The differential subthreshold swing, defined as $S.S. = d(V_{\mathrm{g}})/d(\log_{10} J)$ and normalized by the Boltzmann tyranny, is also plotted. The pristine structure exhibits the widest range ing within the classical limit ($S.S./B.T. = 1$). In contrast, the ranges for the 6-row and 8-row structures are notably reduced, though they preserve partial linearity, whereas the 4-row structure shows only a very limited sloped region. Quantitatively, the minimum normalized values are $S.S./B.T._\text{min} = 0.96$ (6 rows), 0.85 (8 rows), and 0.84 (4 rows), all surpassing the classical limit of $S.S./B.T. = 1$.
}
\label{fig:SS_J_Compare}
\end{figure*}

Using the transmission spectra of each structure, the current density $J$ (hereafter referred to simply as the current) is calculated from the Landauer formula [Eq.~\ref{eq:Landuaer2}] at room temperature ($T = 300$ K) and for various gate voltages $V_{\mathrm{g}}$, under a fixed source–drain bias of $0.05$ V for all configurations. The resulting current and the corresponding subthreshold swing as functions of gate voltage are shown in Figure~\ref{fig:SS_J_Compare}, with the classical limit—Boltzmann Tyranny (B.T.)—indicated for reference.

In the room-temperature $\log_{10}(J)–V_g$ curves, the pristine structure maintains its subthreshold swing (S.S.) slightly below the classical limit ($S.S./\text{Boltzmann tyranny} = 1$) over the widest range. By contrast, the substitutionally doped superlattices (4-row, 6-row, and 8-row) exhibit reduced linear regions, with the 8-row and 6-row configurations retaining broader ranges than the 4-row case. This trend can be understood from the band structure analysis: the pristine structure exhibits the largest band gap ($E_{\mathrm{g}}$), followed by the 8-row and 6-row configurations, while the 4-row configuration has the smallest $E_{\mathrm{g}}$. According to fundamental semiconductor theory, the band gap exerts an exponential influence on the density of thermally excited carriers; a smaller bandgap $E_{\mathrm{g}}$ yields a higher carrier concentration, making it more difficult for the gate to suppress the current and thereby increasing $J_{\mathrm{OFF}}$. Consistent with this reasoning, the expected order of $J_{\mathrm{OFF}}$ is 4-row $>$ 6-row $>$ 8-row $>$ pristine. Examination of the minimum current values in the room-temperature $\log_{10}(J)–V_g$ curves confirms this trend.

\begin{figure*} [ht] 
\centering
\includegraphics[width=\textwidth]{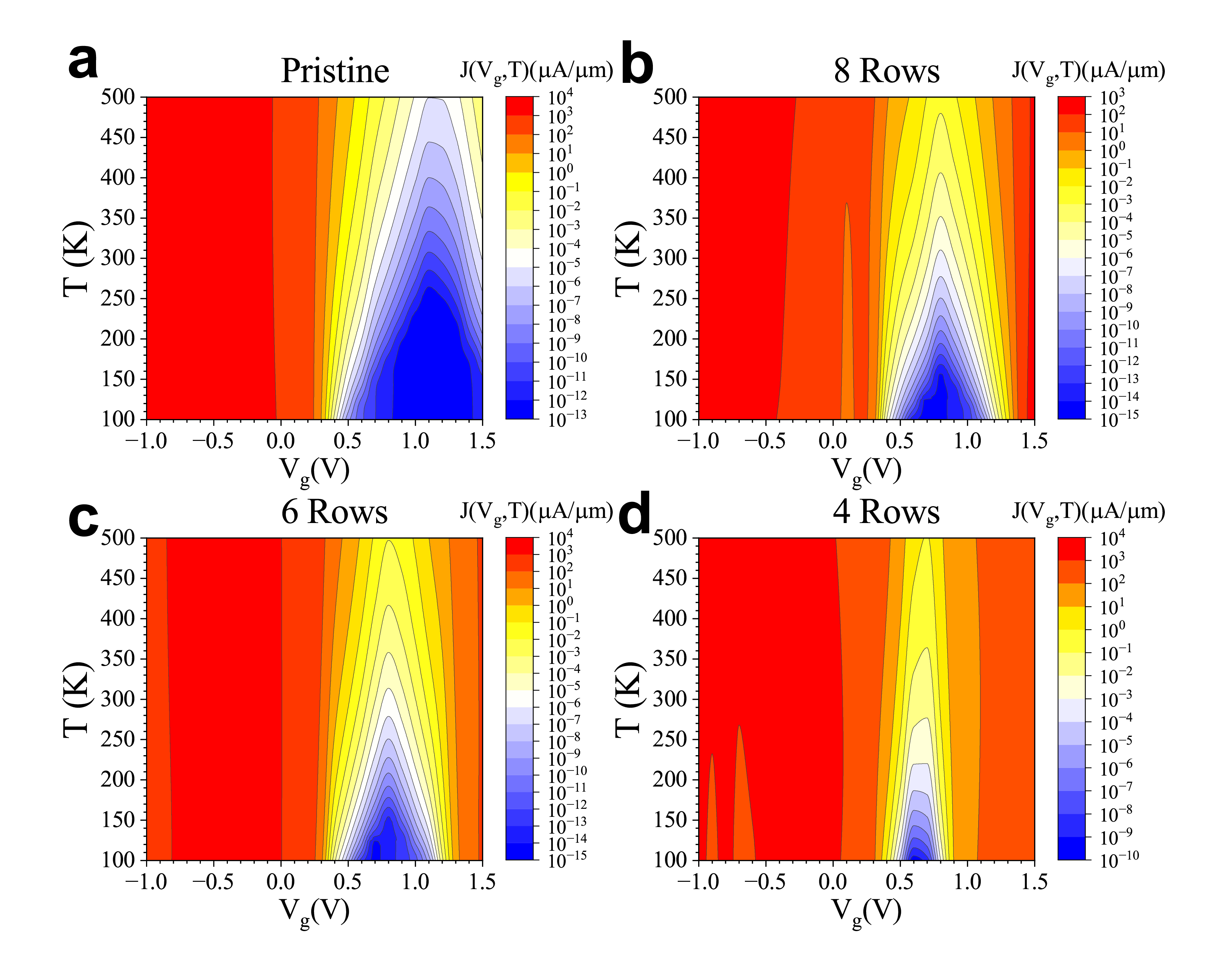}
\caption{
Contour plots of the current densities $J(V_g, T)$ for the four FET structures over gate voltage range from $-1$ to $1.5$ V and the temperature range 100–500 K, with the source–drain bias fixed at 0.05 V. As temperature increases, the current at a given $V_g$ rises, leading to a progressive shrinkage of the low-current (blue) region, which corresponds to the OFF state.
}
\label{fig:J_contour}
\end{figure*}

If these transistors are to be employed as electronic components in future integrated circuits, they will inevitably operate under elevated temperatures. Consequently, understanding temperature effects is of critical importance for assessing the performance of such nano-FETs. To investigate this influence of temperature, we analyzed the full current density spectra, $J(V_{\mathrm{g}}, T)$, for the four FET structures depicted in Figure~\ref{fig:scheme}$\bf{a}$ and present them as contour plots over gate voltage $V_{\mathrm{g}}$ and temperature $T$, as shown in Figure~\ref{fig:J_contour}. The temperature parameter is expanded from room temperature to a range of 100 K to 500 K in the Fermi–Dirac distribution functions of the Landauer formula. The gate voltage utilized for current modulation ranges from $-1$ to $1.5$ V.

From the contour plots of $J(V_g, T)$, a common trend is observed across all four FET structures: as temperature increases, the current at a given $V_g$ gradually rises in the insulating regime, leading to a shrinkage of the low-current region (blue area in the plots, corresponding to the OFF state). This behavior originates from the thermal broadening of the Fermi–Dirac distribution. At high temperatures, the difference between the left and right lead distributions produces a broad and shallow integration window, averaging over a wide energy range in the transmission. Because the nitrogen-doped superlattices have reduced band gaps, the calculated OFF-state current remains significantly influenced by thermionic leakage. In contrast, at low temperatures, the Fermi–Dirac distribution becomes much sharper—approaching a Dirac delta function—which more faithfully follows the transmission spectrum, thereby producing a well-defined OFF region in the current density.

\begin{figure*} [ht] 
\centering
\includegraphics[width=0.7\textwidth]{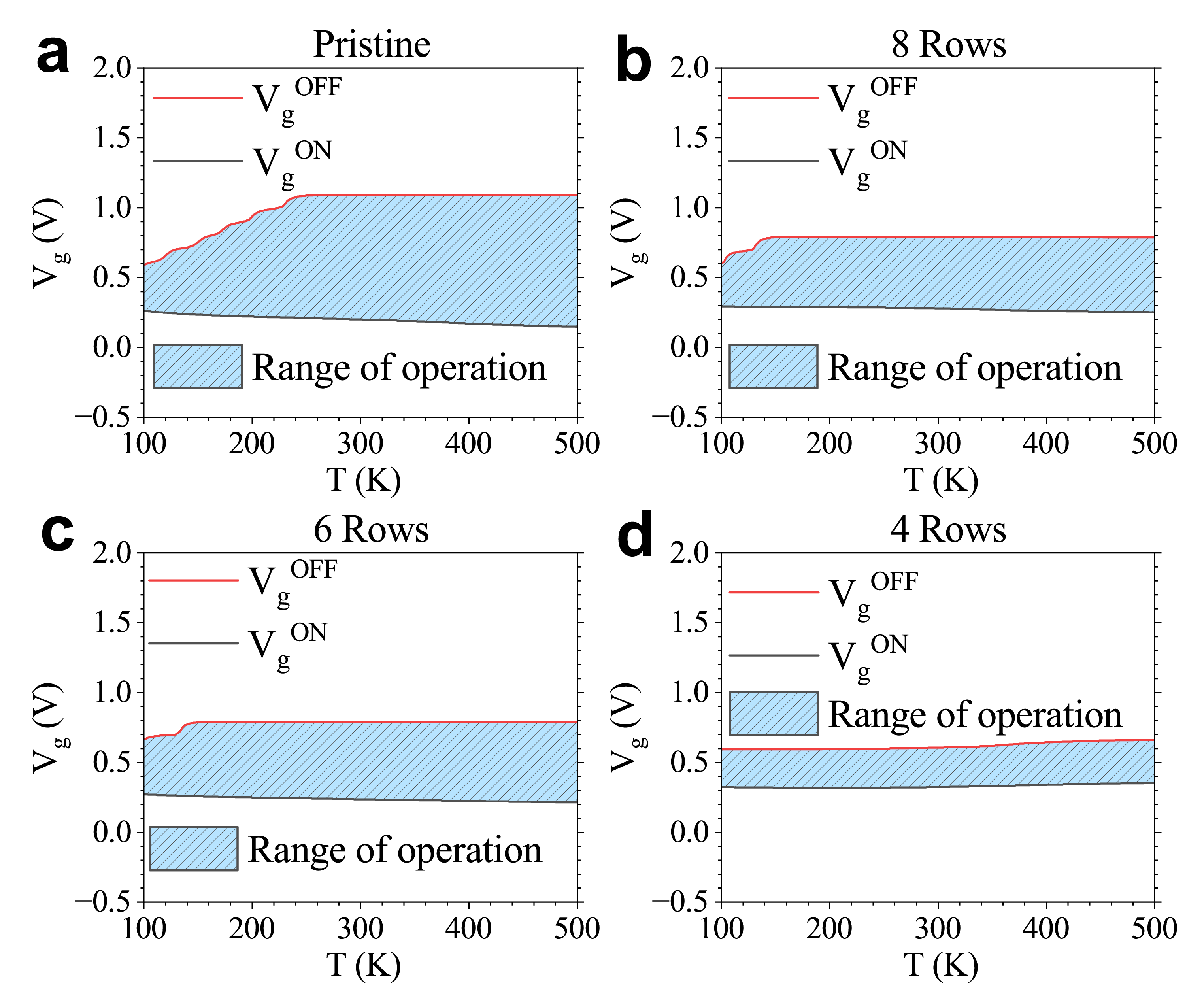}
\caption{
Temperature dependence of the optimal operating range for the four FET structures. When the temperature exceeds approximately 250 K, the upper and lower bounds, $V_g^\mathrm{ON}$ and $V_g^\mathrm{OFF}$, become nearly temperature-independent, and the operating window remains essentially constant. The magnitudes follow the order: pristine ($\sim$1.0 V) $>$ 8-row $\approx$ 6-row ($\sim$0.5 V) $>$ 4-row ($\sim$0.25 V). Furthermore, the $V_g^\mathrm{OFF}$ of the pristine structure ($\sim$1.1 V) is the largest among the four, followed by the 8-row and 6-row structures ($\sim$0.75 V), while the 4-row configuration exhibits the smallest value ($\sim$0.6 V).
} 
\label{fig:OPrange}
\end{figure*}

From the $S.S./\text{Boltzmann Tyranny}$ analysis in Figure~\ref{fig:SS_J_Compare}, we define the optimal operating range, $(V_g^{\mathrm{OFF}},V_g^{\mathrm{ON}})$, offering deeper insight into how the switching behavior of the FETs evolves with temperature. The optimal switching voltages $V_g^{\mathrm{ON}}$ and $V_g^{\mathrm{OFF}}$ are defined by the criterion $S.S./\text{Boltzmann Tyranny} \leq 3$, which delineates the linear region of the $\log_{10}(J)$–$V_g$ curve; the boundaries of this region correspond to the current densities of the ON and OFF states. The optimal operating range, which is dependent on temperature, is illustrated by the interval between $ V_g^{\mathrm{OFF}}$ and $ V_g^{\mathrm{ON}}$ (highlighted in the blue area), as shown in Figure~\ref{fig:OPrange} for four FET structures. The results indicate that both $V_g^{\mathrm{ON}}$ and $V_g^{\mathrm{OFF}}$ are largely insensitive to temperature, with only minor shifts observed below $\sim$250 K, demonstrating the thermal stability of the switching voltages across all structures. A comparison of the operating ranges yields the following order: pristine ($\sim$1.0 V) $>$ 8-row $\approx$ 6-row ($\sim$0.5 V) $>$ 4-row ($\sim$0.25 V). This trend is consistent with the relative widths of the linear regions in the room-temperature $\log_{10}(J)$–$V_g$ characteristics and reflects the same underlying physical mechanism. Consequently, the 4-row structure is considered unsuitable for practical FET applications due to its excessively narrow operating range. A closer inspection of the absolute values of $V_g^\mathrm{OFF}$ reveals that the pristine structure exhibits the largest value ($\sim$1.1 V), followed by the 8-row and 6-row structures ($\sim$0.75 V), while the 4-row structure shows the smallest ($\sim$0.6 V). In modern chip design, however, gate voltages exceeding $\sim$1 V are generally undesirable, as the resulting large vertical electric field across the oxide can induce and accumulate defects, ultimately causing a permanent loss of insulation known as time-dependent dielectric breakdown (TDDB). Thus, although the pristine structure provides the widest operation range, its excessively high $V_g^\mathrm{OFF}$ limits its practical applicability. By contrast, the 6-row and 8-row structures strike a more favorable balance between operation range and threshold voltages, making them more promising candidates for FET applications in future integrated circuits.

\begin{figure*} [ht] 
\centering
\includegraphics[width=1.0\textwidth]{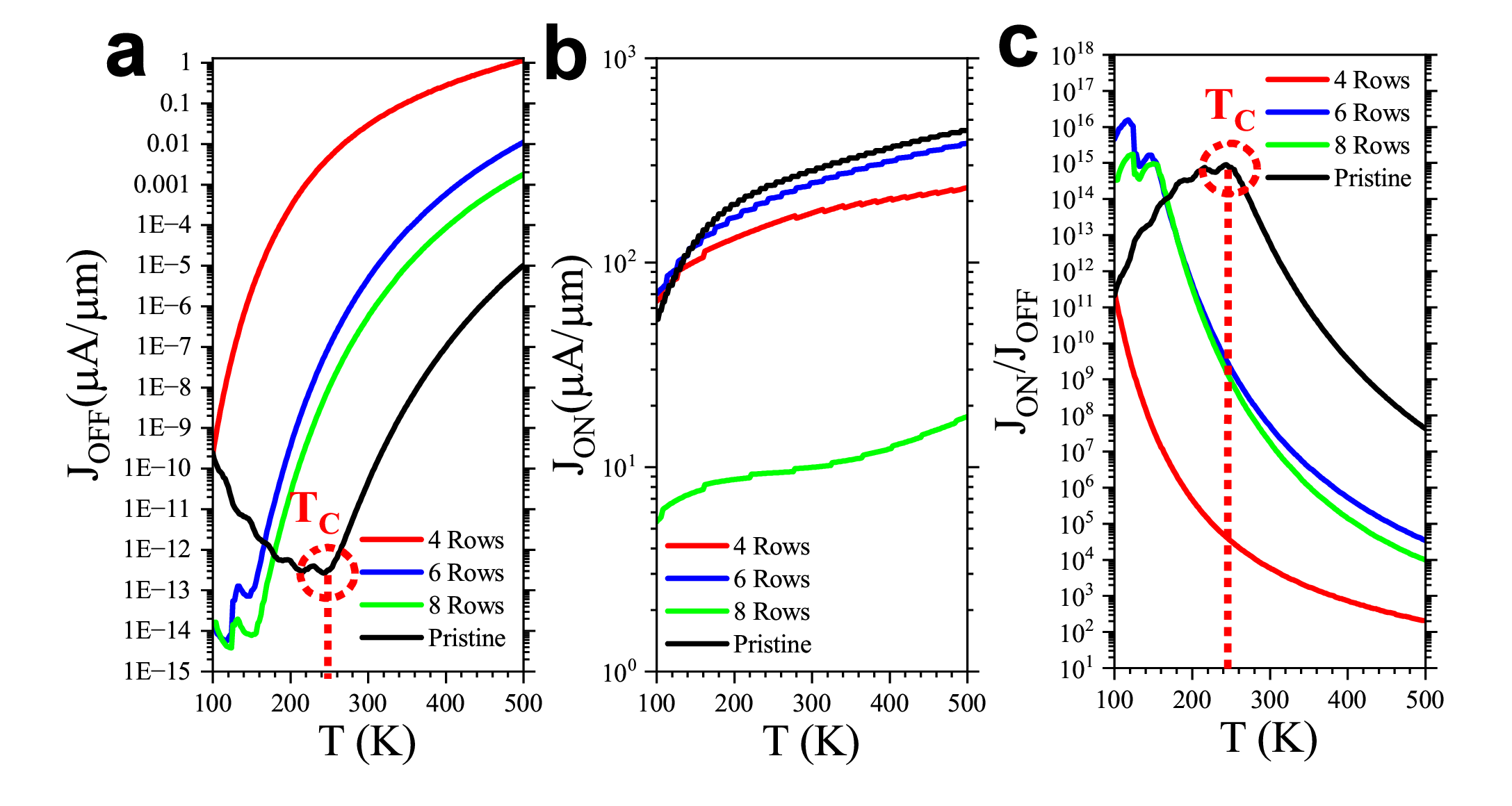}
\caption{
Temperature dependence of a) $J_{\mathrm{OFF}}$, b) $J_{\mathrm{ON}}$, and c) the ON/OFF ratio $J_{\mathrm{ON}}/J_{\mathrm{OFF}}$ for the four FET structures. All structures exhibit the same general trend: at sufficiently high temperatures (above $\sim$250 K), $J_{\mathrm{OFF}}$ increases with temperature, while $J_{\mathrm{ON}}/J_{\mathrm{OFF}}$ decreases. In this regime, the ordering of $J_{\mathrm{OFF}}$ is pristine $<$ 8-row $<$ 6-row $<$ 4-row, whereas the ordering of $J_{\mathrm{ON}}/J_{\mathrm{OFF}}$ is pristine $>$ 6-row $\approx$ 8-row $>$ 4-row.
}
\label{fig:J_ON_OFF}
\end{figure*}

The voltages $V_g^{\mathrm{ON}}$ and $V_g^{\mathrm{OFF}}$ correspond to the ON current $J_{\mathrm{ON}}$ and OFF current $J_{\mathrm{OFF}}$, respectively, as shown in Figure~\ref{fig:J_ON_OFF}$\bf{a}$ and $\bf{b}$, together with the ON/OFF ratio $J_{\mathrm{ON}}/J_{\mathrm{OFF}}$ in panel $\bf{c}$. A higher ON/OFF ratio is generally preferred, as it reflects superior switching performance and suitability for practical chip design. Panel $\bf{c}$ demonstrates that the pristine channel (black solid line) exhibits a higher ON/OFF ratio than all nitrogen-doped superlattice configurations at temperatures above $\sim$170 K. However, around 270 K, the ON/OFF ratio of the pristine structure decreases and falls below that of the doped structures. This trend arises primarily from the behavior of the OFF current in panel $\bf{a}$, since all configurations exhibit comparable ON currents in panel $\bf{b}$, differing by only one to two orders of magnitude.

The OFF current of the FET with a pristine WSe$_2$ channel defines a critical temperature $T_c$, at which the device can be turned off most effectively. Based on this criterion, the temperature dependence can be divided into two regimes relative to $T_c$:~\cite{YC_FET_2025}
\begin{enumerate} [label=(\roman*)]
\item[(i)] The OFF current asymptotically approaches the thermionic emission current, which increases exponentially with temperature and depends only on the barrier height in the scattering region—independent of barrier width—consistent with Richardson’s law [Eq.~\ref{eq:Richadson}].
\item[(ii)]  Low-temperature regime ($T < T_c$): dominated by quantum tunneling current, where $J \approx G_0 \tau[\mu(V_g)]V_{ds}$.
\end{enumerate}
As temperature decreases, the Fermi–Dirac broadening window (given by the difference between the left and right lead distributions) sharpens, reducing the linear operation range. Consequently, the tail of the linear region increases, leading to higher tunneling current. This behavior is consistent with Figure~\ref{fig:OPrange}$\bf{a}$, where the operation range contracts and $V_g^{\mathrm{OFF}}$ shifts to more negative values, corresponding to increased current. Thus, in the pristine case, the OFF current is dominated by thermionic leakage at high temperatures and by tunneling at low temperatures, resulting in a degraded ON/OFF ratio at low $T$.

By contrast, the nitrogen-doped superlattices are dominated by thermionic emission leakage across the entire temperature range considered. Since thermionic current is negligible at low temperatures, these doped configurations maintain excellent ON/OFF ratios that surpass the pristine case at low $T$. In other words, nitrogen doping effectively suppresses quantum tunneling current. Notably, the crossover temperature $T_c$ increases as channel length decreases, reaching room temperature at $\sim$10 nm. Given that channel length scaling is presently limited to $\sim$10 nm due to unsuppressable quantum tunneling, nitrogen-doped configurations may provide a promising route to overcome this fundamental scaling barrier.

\section{Conclusion}\label{sec:conclusion}

We conducted first-principles computations to examine the electronic structures of both pristine WSe$_2$ monolayers and superlattices of WSe$_2$ monolayers with periodic nitrogen substitution. Unlike random doping, which frequently introduces in-gap impurity states, periodic nitrogen doping primarily tunes the band gap. Periodic nitrogen substitution tunes the electronic structures in a clean, crystalline manner—shrinking the gap in the order pristine $>$ 8-row $>$ 6-row $>$ 4-row, which directly impacts the switching characteristics of the FETs.

Applying NEGF-DFT with a realistic gate architecture, we calculated the transmission spectra of the four structures and observed that the doped configurations exhibit reduced transmission compared to the pristine case. The superlattice potential generated by the N rows partitions the channel into quantum wells, producing periodic barriers that reshape $\tau(E)$ (notably a steeper onset near the VBM and side peaks above the CBM) and thereby modulate current under gating. 

Electron transport analysis employing the Landauer formalism determines the dependency of the current density $J(T,V_g)$ on the temperature range (100 K, 500 K) and the voltage range (-1 V, 1.5 V). Regarding device performance, we focused on the practically relevant temperature range above 250 K (up to 500 K). The results reveal common trends among the four structures:$J_\mathrm{OFF}$ increases with temperature, while the ON/OFF ratio $J_\mathrm{ON}$/$J_\mathrm{OFF}$ decreases, indicating that elevated temperature degrades switching performance. Nevertheless, $V_g^\mathrm{ON}$, $V_g^\mathrm{OFF}$, and the operation range $\Delta V_g$ remainslargely insensitive to temperature, demonstrating good thermal stability of the threshold voltages. A comparative analysis further shows that the 4-row structure, with excessively large $J_\mathrm{OFF}$, extremely low ON/OFF ratio, and narrow operation range, is unsuitable for practical FET applications. The pristine structure exhibits the best performance across all metrics, but its high $V_g^\mathrm{OFF}$ ($\sim$1.1 V) makes it less practical, as such a large threshold voltage could possibly induce time-dependent dielectric breakdown (TDDB) of the oxide layer, thereby limiting device reliability. In contrast, the 6-row and 8-row structures, while slightly inferior to pristine in terms of performance, exhibit more favorable $V_g^\mathrm{OFF}$ values ($\sim$0.75 V), achieving a balance between reasonable threshold voltage and stable operation range, making them more promising candidates for future FET integration.

In summary, this study delivers a comprehensive picture of periodic nitrogen substitution in monolayer WSe$2$ FETs, connecting band-gap engineering to gate-controlled transport and device performance. We elucidated the impact of periodic doping on the band structure, transmission characteristics, and electrostatic potential distribution, while also contrasting their disparities in critical performance metrics—subthreshold swing, OFF-state current density, ON/OFF current ratio, and operational range—and assessed their temperature dependence. These results offer important guidelines for the design of high-performance and thermally stable two-dimensional FETs, highlighting their potential value in next-generation nanoelectronics.


\begin{acknowledgments}
The authors thank MOE ATU, NCHC,  National Center for Theoretical Sciences(South), and NSTC (Taiwan) for support under Grant NSTC 111-2112-M-A49-032-. also supported by NSTC T-Star Center Project: Future Semiconductor Technology Research Center under NSTC 114-2634-F-A49-001-. This work was financially supported under Grant No. NSTC-113-2112-M-A49-037-, and supported by NSTC T-Star Center Project: Future Semiconductor Technology Research Center under NSTC 114-2634-F-A49-001-, and also supported in part by the Ministry of Science and Technology, Taiwan. We thank to National Center for High-performance Computing (NCHC) for providing computational and storage resources.
\end{acknowledgments}





\bibliography{ref}


\end{document}